\documentclass[12pt]{article}
\input{epsf}
\def\@fmsl@sh#1#2#3{\m@th\ooalign{$\hfil#1\mkern#2/\hfil$\crcr$#1#3$}}
 \def\eq#1\en{\begin{equation}#1\end{equation}}
\def\s[#1,#2]{[#1\stackrel{\star}{,}#2]}
\def\sx[#1,#2]{[#1\stackrel{\star_{x}}{,}#2]}

\newcommand{\nc}{\newcommand}
\nc{\beq}{\begin{equation}}
\nc{\eeq}{\end{equation}}
\nc{\beqa}{\begin{eqnarray}}
\nc{\eeqa}{\end{eqnarray}}

\def\bc{\begin{center}}
\def\ec{\end{center}}

\def\gsim{\mathrel{\mathpalette\atversim>}}

\def\bc{\begin{center}}
\def\ec{\end{center}}

\def\gsim{\mathrel{\rlap{\lower4pt\hbox{\hskip1pt$\sim$}}

    \raise1pt\hbox{$>$}}}       

\def\gsim{\mathrel{\rlap{\lower4pt\hbox{\hskip1pt$\sim$}}
    \raise1pt\hbox{$>$}}}       

\begin{document}
\makeatletter
\def\fmslash{\@ifnextchar[{\fmsl@sh}{\fmsl@sh[0mu]}}
\def\fmsl@sh[#1]#2{%
  \mathchoice
    {\@fmsl@sh\displaystyle{#1}{#2}}%
    {\@fmsl@sh\textstyle{#1}{#2}}%
    {\@fmsl@sh\scriptstyle{#1}{#2}}%
    {\@fmsl@sh\scriptscriptstyle{#1}{#2}}}
\def\@fmsl@sh#1#2#3{\m@th\ooalign{$\hfil#1\mkern#2/\hfil$\crcr$#1#3$}}
\makeatother

\thispagestyle{empty}
\begin{titlepage}
\boldmath
\begin{center}
  \Large {\bf On a Formulation of Qubits in Quantum Field Theory}
    \end{center}
\unboldmath
\vspace{0.2cm}
\begin{center}
{ 
{\large Jacques Calmet$^{a}$}\footnote{calmet@ira.uka.de} {\large and} 
{\large Xavier Calmet$^{b}$}\footnote{x.calmet@sussex.ac.uk}
}
 \end{center}
\begin{center}
$^{a}${\sl Karlsruhe Institute of Technology (KIT), Institute for Cryptography and Security, \\
Am Fasanengarten 5, 76131 Karlsruhe, Germany
}
\\
$^{b}${\sl Physics and Astronomy, 
University of Sussex,   Falmer, Brighton, BN1 9QH, UK 
}
\end{center}
\vspace{\fill}
\begin{abstract}
\noindent
Qubits have been designed in the framework of quantum mechanics. Attempts to formulate the problem in the language of quantum field theory have been proposed already.
In this short note we refine the meaning of qubits within the framework of quantum field theory. We show that the notion of gauge invariance naturally leads to a generalization of qubits to QFTbits which are then the fundamental carriers of information from the quantum field theoretical point of view. The goal of this note is to stress the availability of such a  generalized concept of QFTbits. 
\end{abstract}  
\end{titlepage}



\newpage

Nowadays quantum computing and information theory are based on the notion of qubits \cite{book} which are the basic building blocks in these fields when considered solely in the framework of quantum mechanics.
Two possible states for a qubit are the states  $|0\rangle$ and $|1\rangle$. From a quantum mechanical point of view, this is a valid description of an electron with spin down or up. A physical state   $|\psi\rangle$ can be a superposition of the states $|0\rangle$ and $|1\rangle$: $|\psi\rangle=a|0\rangle+b|1\rangle$ with $\sqrt{a^2+b^2}=1$.
This description is fully satisfactory from a quantum mechanical point of view, however it is now well understood that quantum mechanics needs to be extended to quantum field theory (QFT) to acknowledge our understanding of the physical world. We will argue that the notion of  qubits needs to be adapted to be physical objects within the framework of quantum field theory.

 In this very short note we propose a generalization of qubits within the realm of QFT. The main goal of this note is to stress the existence of this generalized concept of QFTbit.
The motivation for this extension comes from well known facts in QFT. First, 
quantum mechanics needs to be extended to quantum field theories since the former can only describes processes where the number of particles is conserved while quantum field theories are able to describe processes where particles are created or annihilated. In physical processes, the number of particles is in general not conserved since particles can decay or annihilate. 
A second motivation to extend the notion of qubits to QFTbits comes from the fact that interactions, in modern models of the world based on quantum field theories, are implemented via gauge interactions. A physical observable has to be gauge invariant. In that sense an electron, which can be modeled by a qubit, is never observed alone, but with a cloud of photons and electron/position pairs fluctuating around it.
To require gauge invariance naturally leads to a generalization of qubits to QFTbits which are then the fundamental carriers of information from the quantum field theoretical point of view.  Our results imply an extra layer of complexity within new approaches to quantum information.  A nice investigation of the formulation of qubit into quantum electrodynamics has been proposed in  \cite{BB}. However, a consequence of this formulation looks to have been omitted so far: gauge invariance.

As mentioned already, the major difference between quantum mechanics and quantum field theory is that the number of particles or states is not fixed in the latter one. Within quantum field theories, particles are created and annihilated constantly. The generalization of the quantum mechanical treatment of electrons interacting with photons leads to quantum electrodynamics (QED). Within QED it makes little sense to talk about an isolated electron with spin up or down since an electron will always emit photons that may be even too soft to observe. This leads to the well-known phenomenon of infrared divergencies in QED (see e.g. \cite{Weinberg:1995mt}).  QED is described by the following action 
\begin{eqnarray}
S=\int d^4x (\bar \psi(x) (\gamma^\mu i D_\mu- m) \psi(x) -\frac{1}{4} F^{\mu\nu}(x)F_{\mu\nu}(x))
\end{eqnarray}
 where $D_\mu=\partial_\mu - i e A_\mu(x)$ is the covariant derivative, $e$ is the electric charge, $F_{\mu\nu}(x)=\partial_\mu A_\nu(x) - \partial_\nu A_\mu(x)$, $\gamma_\mu$ are the Dirac matrices,  $\psi(x)$ is the quantum field which describe the electron field, $\bar \psi(x) = \psi^\dagger(x) \gamma_0$ and $A_\mu(x)$ is the electromagnetic potential which is related to the electric and magnetic fields.  
 Another property of QED is that  the action  describing this quantum field theory is invariant under local U(1) gauge transformations
 \begin{eqnarray}
 \psi^\prime(x)&=& \exp(i \alpha(x)) \psi(x) \\
 A_\mu^\prime(x)&=&  A_\mu(x) -\frac{1}{e}  \partial_\mu \alpha(x). 
 \end{eqnarray}
 On the other hand, the field which represents the electron  and the positron is not gauge invariant. However, only gauge invariant quantities are observable. This was a motivation for Dirac to formulate QED using dressed fields \cite{Dirac:1955uv}, see also \cite{Bagan:1999jf} for more recent work in that direction where the connection to the infrared problem was studied. Dirac introduced:
\begin{eqnarray} \label{dressed}
\psi_f(x) \equiv \exp \left ( - i e \int d^4z f^\mu(x-z) A_\mu(z) \right ) \psi(x)
\end{eqnarray}
where $A_\mu$ is the electromagnetic four-vector which represents the photon field.
Note that $\psi_f(x) $ is locally  but not globally gauge invariant as long as $\partial_\mu f^\mu(z)=\delta^{(4)}(z)$. The spin of $\psi_f(x)$ can clearly be up or down as it was before for the qubits and it is determined by the spin of $\psi(x)$ which is the only quantity that carries a spinor index. A specific function $f^\mu(x-z)$ was proposed by Dirac:
\begin{eqnarray} 
\psi_D(x) \equiv \exp \left ( - i e \frac{\partial^i A_i(x)}{\nabla} \right ) \psi(x).
\end{eqnarray}
This representation shows clearly that the electron is accompanied by an electric field and the object $\psi_D(x)$ is non-local.
 
It is thus natural to propose to extend the notion of qubits to that of QFTbits which we define as the states
\begin{eqnarray}
|0_f(x)\rangle \equiv \exp \left ( - i e \int d^4z f^\mu(x-z) A_\mu(z) \right ) |0(x)\rangle
\end{eqnarray}
and
\begin{eqnarray}
|1_f(x)\rangle \equiv \exp \left ( - i e \int d^4z f^\mu(x-z) A_\mu(z) \right ) |1(x)\rangle,
\end{eqnarray}
which are parametrized by the function $f$ as was $\psi_f(x)$ in eq. (\ref{dressed}). The states $|0_f(x)\rangle$ and $|1_f(x)\rangle$ defined this way are locally gauge invariant as long as  $\partial_\mu f^\mu(z)=\delta^{(4)}(z)$. For example, they can thus be used to describe an electron with spin up or down and the electromagnetic  field that surrounds it. It is trivial to obtain the momentum space representation by doing a Fourier transformation on both sides of these equations.  This generalization based on the technique of field dressing is similar to that proposed in \cite{LMM} which relies heavily on the work of  \cite{Bagan:1999jf} who have chosen a specific function $f^\mu(x-z)$ in equation (\ref{dressed}). For different physical problems, scattering problems, bound state problems etc, the choice of asymptotic states can be different and it is best to keep an open mind concerning the choice of the dressing function  $f$. Again, the main message is that a qubit is never a free field from a quantum field theoretical point of view, but  it is always dressed with photons and a cloud of electron positron pairs which can freely emerge from the vacuum. 

As before we can obtain a superposition by taking
\begin{eqnarray}
|\psi_f\rangle=a|0_f\rangle+b|1_f\rangle
\end{eqnarray}
with $\sqrt{a^2+b^2}=1$.
Another similar approach would be to search for classical solutions of QED and use them as representation of QFTbits. An entangled state can be obtained easily:
\begin{eqnarray}
|0_f(x) 0_f(y) \rangle \equiv \exp \left ( - i e \int d^4z f^\mu(x-z) A_\mu(z) \right ) \exp \left ( - i e \int d^4z^\prime f^\nu(y-z^\prime) A_\nu(z^\prime) \right ) |0(x) 0(y)\rangle
\end{eqnarray}

Although we have focussed here on electrons as qubits, the very same reasoning applies to any particle used as a qubit, whether photons or effective two dimensional system such as atoms or nuclei. 
This is why it is possible to speak of a generalization of the concept of qubits. In the case of photons, one can define a gauge invariant four-potential \cite{Calmet:2010cb}
\begin{eqnarray} 
A_{\mu}  = \frac{i}{2 e}  \Omega^\dagger \stackrel{\leftrightarrow}
{D}_\mu \Omega.
\end{eqnarray}
with 
\begin{eqnarray} 
\Omega = \exp \left ( i e \sigma(x)  \right ), 
 \end{eqnarray}
with $\delta \sigma(x) = - \alpha(x)/e$ under a U(1) gauge transformation. Alternatively, one can use the field strength tensor $F^{\mu\nu}$ to describe the photon in a gauge invariant way. With this new gauge invariant formulation, it becomes indeed possible to define different species of quantum information. One is still a system described by quantum mechanics as in nowadays existing quantum information. Other would consist in selecting a framework based on quantum field theory which is important when bound states of leptons or more complex QED systems are considered. Indeed,  besides the foundational distinction between qubits and QFTbits, the difference between the two approaches becomes important once interactions are considered between physical states as for example done in \cite{BB}. 

Quantum information relies on the notion of entanglement of quantum states. Indeed, entanglement enables operations, which are impossible classically. This remark covers the well-known discussion related to the violation of Bell's inequality.
Thirring et al. \cite{Thirring}  discuss nicely how the concept of entanglement changes with respect to the possible factorizations of the total algebra, which describes the quantum states. They apply their discussion to the familiar case of qubits. Another recent algebraic approach \cite{Borsten:2009ae} introduces superqubits as a supersymmetric generalization of qubits and outlines the superentanglement of two and three cubits. We do not discuss in this note the in-depth consequence of the introduction of QFTbit. However, entanglement is a meaningful concept in our formulation as briefly outlined in this note.
An important remark is that since we do not rely on a specific algebra  to construct our states, the first of the two misconceptions in current relativistic quantum information pointed out by Br\` adler \cite{Bradler}  is not relevant for our approach. We avoid the second conceptual misconception related to the investigation of channel capacities since we have a classical relativistic framework.

We foresee different implications of our considerations. For example, the second quantification of the quantum Byzantine agreement  \cite{maurer}  would change the problem quite a bit. Indeed at the quantum field theoretical level, electromagnetism develops some non-linear terms  \cite{Heisenberg:1935qt} due to loop corrections involving electrons. One could use  the non-linear interactions between photons (communication channels) sent by the generals and the general-in-chief to solve the Byzantine problem. Strictly  speaking it is not possible to reduce the problem to three generals as done in  \cite{maurer}  since quantum field theoretical effects will open up all the channels. There are also interesting questions raised for quantum cryptography. One could try to use the phase of the spinors that depends on the electromagnetic potential, which is a second quantification effect, for secured communications. We stress that although quantum mechanics remains linear in the wave function, the action and thus the Hamiltonian can be non-linear. This  is the case for all forces of nature known so far at least at the quantum field theoretical level. These non-linear effects are usually disregarded in quantum information theory. Their effects need to be scrutinized in details.

These are just a couple of potential applications of our idea. The main message is that at the quantum field theoretical level, it is impossible to consider a given particle on its own. If electrons are present, photons are present and vice versa. Once again, we do not claim that there is anything wrong with the usual formulation of quantum information theory. We suggest to go beyond it. The standard formulation of quantum information does not usually imply formulating an action, or equivalently a Hamiltonian, but rather focuses on a few solutions (i.e the qubits). We have shown that  the spectrum of solutions of the underlying quantum field theory is richer.

Finally let us stress  that the formalism we propose here is well-suited for quantum information theory and quantum computing. Indeed  the non-local effects we take into account  here will have important implications in these areas.

\bigskip



\bigskip


\end{document}